\documentclass[a4paper,11pt]{article}
\usepackage{pos}

\newcommand{\beq}{\begin{eqnarray}}
\newcommand{\eeq}{\end{eqnarray}}

\usepackage{bbm}

\title{Speed of sound exceeding the conformal bound in dense 2-color QCD}

\author*[a,b]{Etsuko Itou}
\author[c]{Kei Iida}

\affiliation[a]{Yukawa Institute for Theoretical Physics, Kyoto University, Kitashirakawa Oiwakecho, Sakyo-ku, Kyoto 606-8502, Japan}

\affiliation[b]{Interdisciplinary Theoretical and Mathematical Sciences Program (iTHEMS), RIKEN, Wako, Saitama 351-0198, Japan}

\affiliation[c]{Department of Mathematics and Physics, Kochi University, 2-5-1 Akebono-cho, Kochi 780-8520, Japan}

\emailAdd{itou@yukawa.kyoto-u.ac.jp}
\emailAdd{iida@kochi-u.ac.jp}

\abstract{We review recent works on the Monte Carlo simulations of dense two-color QCD (QC$_2$D) by focusing on the phase diagram, the equation of state, and the sound velocity at nonzero quark chemical potential. A possible upper bound of the sound velocity is known as the conformal bound, namely, $c_s^2/c^2 \leq 1/3$. The sound velocity is below the bound at least in the case of finite-temperature QCD. However, our recent work~\cite{Iida:2022hyy} shows the breaking of this bound in dense QC$_2$D. This phenomenon was previously unknown from any lattice calculations. We also discuss recent related works including lattice studies on QCD at nonzero isospin chemical potential, some effective model analyses, and an analysis based on recent neutron star observations. These works also suggest the breaking of the conformal bound.}

\FullConference{The 40th International Symposium on Lattice Field Theory (Lattice 2023)\\
July 31st - August 4th, 2023\\
Fermi National Accelerator Laboratory\\}

\begin{document}
\maketitle

\section{Introduction}
The conformal bound has been conjectured by A.~Cherman et al.\ in 2009~\cite{Cherman:2009tw} such that {\it{``Sound velocity squared, $c_s^2/c^2 = \partial p /\partial e$, has an upper bound ($1/3$) for a broad class of four-dimensional theories.''}}
Here, $p$ and $e$ denote the pressure and internal energy density in the corresponding thermodynamic system, and $c_s^2/c^2=1/3$ is realized in relativistic free theory. The lattice Monte Carlo results for the sound velocity in finite-temperature QCD satisfy this bound.  Indeed, the sound velocity has a minimum around the pseudo-critical temperature of chiral symmetry breaking and monotonically increases as a function of temperature.  Eventually it approaches $c_s^2/c^2 =1/3$ in the high-temperature limit~\cite{Borsanyi:2013bia, HotQCD:2014kol}.
 \begin{figure}[htbp]
 \begin{center}
    \begin{tabular}{c}
        \includegraphics[keepaspectratio, scale=0.4]{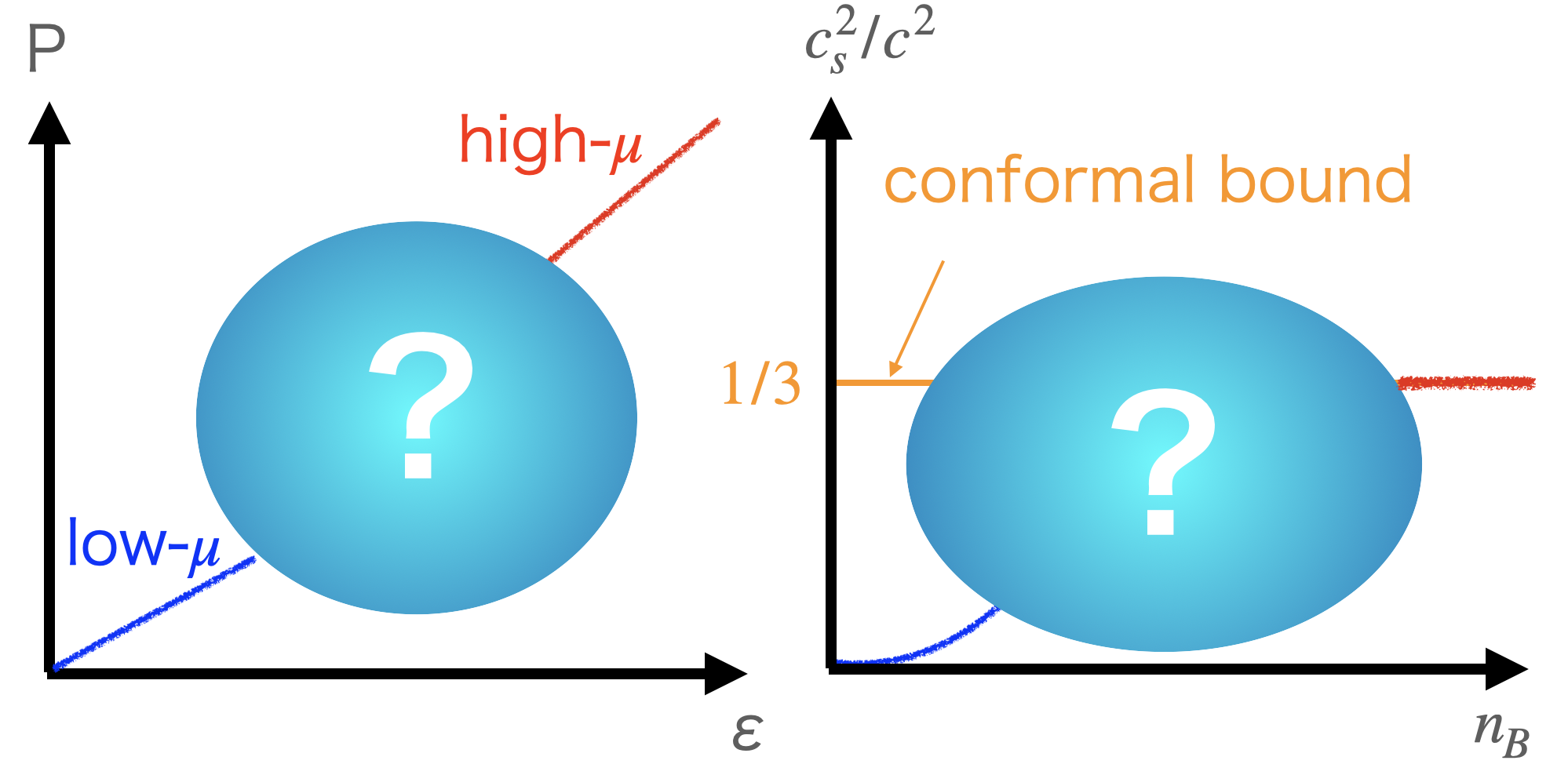}
    \end{tabular}
            \caption{
Schematic pictures of the EoS (left panel) and sound velocity (right panel) for finite-density QCD at zero temperature.
}\label{fig:phase-diagram}
\end{center}
  \end{figure}
On the other hand, the equation of state (EoS) for finite-density QCD at low temperature has not yet been understood by the first-principles calculation because of the notorious sign problem, although the determination of the EoS has been desired since it is related to the mass-radius relation of neutron stars. In both the low-density and high-density limits, we can study the EoS analytically as long as electric charge is ignored.  The theory can be described by the weakly interacting nucleon gas
in a low-density regime, while it approaches perturbative QCD (pQCD) in a high-density regime where the distance between quarks is so small that the coupling is asymptotically free. However, the most interesting density regime, which is related to the physics of neutron stars, is the intermediate density regime where the hadron-quark crossover is expected to occur.

Several works on the phenomenology of neutron stars using observation data~\cite{Masuda:2012ed,Baym:2017whm, Kojo:2021wax} and effective models of dense-QCD studies~\cite{McLerran2019-qh, Kojo2021-mg} suggest that the sound velocity squared at zero-temperature, $c_s^2=\partial p/\partial e$, peaks in the intermediate-density regime.  Furthermore, Kojo and Suenaga~\cite{Kojo2021-wh} argued that a similar peak of $c_s^2$ emerges not only in $3$-color QCD, but also in $2$-color QCD (QC$_2$D).

It is expected that QC$_2$D even at nonzero quark chemical potential could be a good testing ground for understanding dense QCD since QC$_2$D at zero chemical potential shares the same properties as $3$-color QCD, {\it e.g.,} confinement, spontaneous chiral symmetry breaking, and thermodynamic behaviors.
On the other hand, the sign problem is absent in even-flavor dense QC$_2$D  because of the pseudo-reality of fundamental quarks. Based on this motivation, several Monte Carlo studies on dense QC$_2$D have been conducted independently and intensively in recent years (see references in Ref.~\cite{Iida:2022hyy}).
One can conclude that the QC$_2$D phase diagram has been quantitatively clarified; even at fairly high temperatures, $T\approx 100$ MeV, superfluidity can remain~\cite{Ishiguro:2021yxr, Begun:2022bxj}.
Moreover, the EoS and the sound velocity in a low-temperature and high-density regime have been just recently investigated in Ref.~\cite{Iida:2022hyy}. 

In this article, we review recent works on the Monte Carlo simulations of dense QC$_2$D and focus on the sound velocity for QCD(-like) theories by mentioning lattice studies on dense QC$_2$D and QCD at nonzero isospin chemical potential and other works related with the breaking of the conformal bound.

\section{Configuration generation strategy to avoid sign problem and onset problem}
To generate configurations of $3$-color QCD in a low-temperature and high-density regime using the HMC algorithm is accompanied by two serious problems: the sign problem and the onset problem.
The sign problem does not appear if we consider specific kinds of QCD-like theories, for instance, QC$_2$D or QCD with isospin chemical potential.
On the other hand, the onset problem emerges even for such QCD-like theories in the low-temperature and high-density regime where the MC simulation becomes unstable and the calculation does not proceed. Numerically, it is caused by the emergence of near-zero eigenvalues of the Dirac matrix as investigated in Ref.~\cite{Muroya:2000qp}.
Physically, it relates to the spontaneous symmetry breaking (SSB) of U($1$) baryon or isospin. Therefore, one needs to modify the model Lagrangian itself to overcome this problem.
As a standard method for investigating the SSB, we introduce an explicit breaking term of the corresponding symmetry. In QC$_2$D, it can be realized by the diquark source term~\footnote{In the case of QCD at nonzero isospin chemical potential, the corresponding explicit breaking term becomes the pion source term.}, which breaks the symmetry between a baryon (diquark) and an anti-baryon (anti-diquark).
Thus, a possible Lagrangian that helps to study the whole regime of the QC$_2$D phase diagram is given by
\beq
\mathcal{L}= \frac{1}{4} F_{\mu \nu} F_{\mu \nu} + \bar{\psi} (\gamma_\mu D_\mu +m ) \psi + \mu \bar{\psi} \gamma^0 \psi -\frac{j}{2} (\bar{\psi}_1 K \bar{\psi}_2^T -\psi_2^T K \psi_1 ). 
\eeq
Here, $K$ and $j$ are $K=C\gamma_5$ with the charge conjugation operator $C$ and a diquark-source parameter, respectively. 
The indices $1,2$ of $\psi$ denote the flavor label.

On the lattice, the modifications of the fermion part can be implemented by the extended Dirac matrix~\cite{Kogut:2001na, Skullerud:2003yc, Iida:2019rah}.
Utilizing the (naive) Wilson fermion, the fermion action  with the quark number density and diquark source terms can be written by the extended matrix as
\beq
S_F&=& (\bar{\psi_{1}} ~~ \bar{\varphi}) \left( 
\begin{array}{cc}
\Delta(\mu) & J \gamma_5 \\
-J \gamma_5 & \Delta(-\mu) 
\end{array}
\right)
\left( 
\begin{array}{c}
\psi_{1}  \\
\varphi  
\end{array}
\right)
 \equiv  \bar{\Psi} {\mathcal M} \Psi, \nonumber\\ \label{eq:def-M}
\eeq
where
$\bar{\varphi}=-\psi_2^T C \tau_2, ~\varphi=C^{-1} \tau_2 \bar{\psi}_2^T.$
Here, $\Delta(\mu)_{x,y}$  denotes the Wilson-Dirac operator including the number operator and the quark chemical potential $\mu$. 
The additional parameter $J$ corresponds to the diquark source parameter, where  $J=j \kappa$ with the hopping parameter $\kappa$.
The source term allows us to perform the numerical simulation in the superfluid phase.
The Pauli matrix $\tau_2$ in $\bar{\varphi},\varphi$ acts on the color index.
Calculating the extended Dirac matrix ($\mathcal M$) using the HMC algorithm is difficult since it has non-diagonal components, but the square of the matrix can be diagonal. 
On the other hand, $\det[{\mathcal M}^\dag {\mathcal M}]$ corresponds to the fermion action for the four-flavor theory, since a single $\mathcal{M}$ in Eq.\ (\ref{eq:def-M})  represents the fermion kernel of the two-flavor theory.

In our own work, to reduce the number of fermions, we take the root of the extended matrix in the action.
In practice, utilizing the Rational Hybrid Monte Carlo (RHMC) algorithm, we can generate gauge configurations of this action.
We perform the simulation with $(\beta, \kappa,N_s,N_\tau)=(0.80,0.159,16,16)$ where the gauge action is set to the Iwasaki gauge action.
According to Ref.~\cite{Iida:2020emi}, once we introduce the physical scale as $T_c=200$ MeV, where $T_c$ denotes the pseudo-critical temperature of chiral phase transition at $\mu=0$, then our parameter set, $\beta=0.80$ and $N_\tau=16$, corresponds to $a\approx 0.17$ fm and $T\approx 79$ MeV.
The mass of the lightest pseudo-scalar (PS) meson at $\mu=0$ is still heavy in our simulations: $am_{PS}=0.6229(34)$ ($m_{PS}\approx 750 $ MeV).
As for the values of $a\mu$, we generate the configurations at intervals of $a\Delta \mu=0.05$. 
The number of configurations for each parameter is $100$--$300$. 
The statistical errors are estimated by the jackknife method.

\section{Phase structure}
The emergence of the superfluid phase in QC$_2$D has been observed by several independent lattice Monte Carlo studies~\cite{Cotter2012-bh, Cotter2012-zl, Braguta2016-ds, Iida:2019rah, Boz2019-fl, Bornyakov2022-sv, Ishiguro:2021yxr, Buividovich:2021fsa, Buividovich2020-ld}.
Putting all these works together, the superfluid phase appears at least at $T=100$ MeV. This is below $T_c$ at $\mu=0$ but is a higher temperature than naively expected.
It is also reported that even at very high densities ($\mu>1$ GeV), confinement remains~\cite{Ishiguro:2021yxr, Begun:2022bxj}.

According to our works~\cite{Iida:2019rah, Iida:2020emi, Ishiguro:2021yxr, Iida:2022hyy, Itou2018-py}, the phase structure at low temperature can be summarized as in Fig.~\ref{fig:phase-diagram}, where the definition of each phase is shown in Table~\ref{table:phase}. 
 \begin{figure}[htbp]
 \begin{center}
    \begin{tabular}{c}
        \includegraphics[keepaspectratio, scale=0.35]{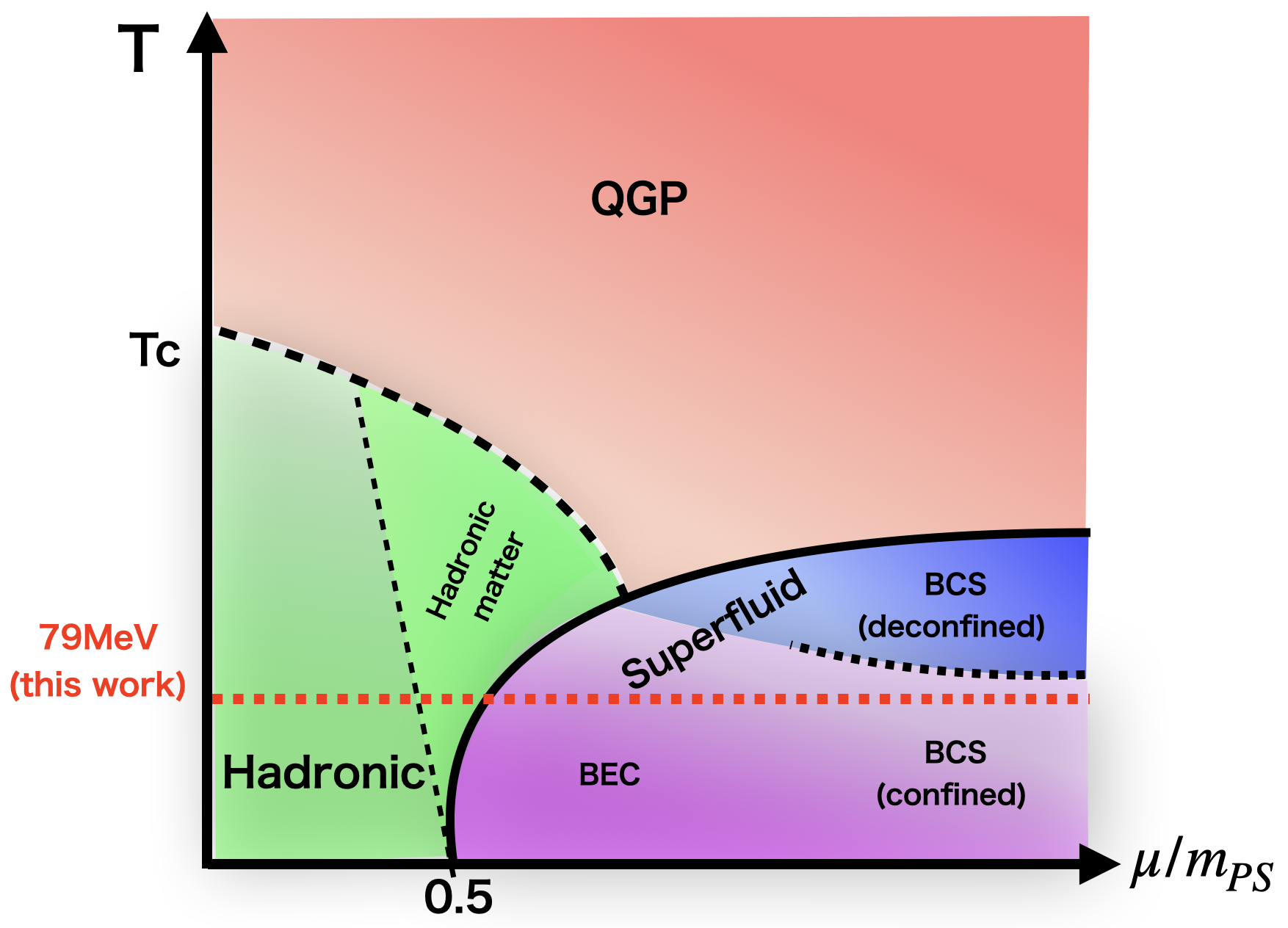}
    \end{tabular}
            \caption{
Schematic 2-color QCD phase diagram.  Each phase is defined in Table~\ref{table:phase}.
}\label{fig:phase-diagram}
\end{center}
  \end{figure}
\begin{table}[h]
\begin{center}
\begin{tabular}{|c||c|c|c|c|}
\hline
 \multicolumn{1}{|c||}{}  & \multicolumn{2}{c|}{Hadronic} &     \multicolumn{2}{c|}{Superfluid}  \\  
\cline{3-3} \cline{4-5}  & & Hadronic matter &  BEC & BCS \\  
 \hline \hline
$\langle |L| \rangle$ & zero  & zero  &    &   \\
$\langle qq \rangle$ & zero  &  zero  & non-zero & non-zero  \\ 
$ \langle n_q \rangle $ &zero &  non-zero & $  0 < \frac{\langle n^{latt.}_q \rangle}{n_q^{\mbox{tree}}} <1 $  & $  \frac{\langle n^{latt.}_q \rangle}{n_q^{\mbox{tree}}} \approx 1 $ \\ 
 \hline
\end{tabular}
\caption{ Definition of phases. } \label{table:phase}
\end{center}
\end{table}
The order parameters that help classify the phases are the Polyakov loop $\langle |L| \rangle$ and diquark condensate $\langle qq \rangle$, whose zero/nonzero values indicate the appearance of confinement and superfluidity, respectively.
We have found that the superfluidity emerges at $\mu_c/m_{PS} \approx 0.5$ as predicted by the chiral perturbation theory (ChPT)~\cite{Kogut2000-so}.
It is natural to use $\mu/m_{PS}$ as a dimensionless parameter of density since the critical value $\mu_c$ can be approximated by $m_{PS}/2$  even if the value of $m_{PS}$ in numerical simulation would be changed~\footnote{In the case of dense $3$-color QCD, the corresponding critical value of $\mu$ would be $\mu_c/m_N \approx 1/3$, where the hadronic-superfluid phase transition is expected to occur. Here, $m_N$ denotes the nucleon mass. }.
We have also confirmed that the scaling law of the order parameter around $\mu_c$ is consistent with the ChPT prediction~\cite{Iida:2019rah}.
Furthermore, we have measured the quark number operator, $n_q^{latt.}\equiv a^3 n_q= \sum_{i} \kappa \langle \bar{\psi}_i (x) (\gamma_0 -\mathbb{I}_4) e^\mu U_{4} (x) \psi_i (x+\hat{4})  + \bar{\psi}_i (x) (\gamma_0 + \mathbb{I}_4) e^{-\mu}U_4^\dag (x-\hat{4} )\psi_i (x-\hat{4})\rangle$.
We have identified the regime where $\langle n^{latt.}_q \rangle$ is consistent with the free quark value $n_q^{\mathrm{tree}}$ (see Eq.~(26) in Ref.~\cite{Hands2006-mh}) as the BCS phase. 
Thus, we have concluded that there are hadronic, hadronic-matter, Bose-Einstein condensed (BEC), and BCS phases at $T=79$ MeV, although there is no clear boundary between the BEC and BCS phases.  Interestingly,
up to $\mu/m_{PS} =1.28 $ ($\mu \lesssim 960$ MeV), the confining behavior remains~\cite{Ishiguro:2021yxr},  while nontrivial instanton configurations have been discovered from calculations of the topological susceptibility~\cite{Iida:2019rah}.
It indicates that a naive perturbative picture, for instance, pQCD, is not yet valid in the density regime studied here.

\section{Calculation strategy: Equation of state and sound velocity at finite $\mu$}
We turn to the EoS at finite density, i.e., the $\mu$-dependence of the internal energy density and pressure.
As for the pressure, in the thermodynamic limit, it satisfies the Gibbs-Duhem relation, so that we can calculate it using
\beq
p (\mu) = \int_{\mu_c}^\mu n_q(\mu') d \mu' . 
\eeq
Note that there is no renormalization for the quark number density as it is a conversed quantity.

Another quantity, which we can calculate easily, is the trace anomaly.
The trace anomaly basically consists of the beta-function for the parameters in the action and the trace part of the energy-momentum tensor. 
Some methods have been proposed to estimate the beta-function or multiplicative renormalization factor at finite $\mu$, for instance, non-perturbative determination of Karsch coefficients~\cite{Karsch:1989fu, Hands:2011ye} or perturbative calculations~\cite{Astrakhantsev:2020tdl}.
Here, we utilize the non-perturbative beta-function for each parameter evaluated at $\mu=T=0$ along the line of constant physics (LCP). Then, the trace anomaly is explicitly given by
\beq
e-3p &=& \frac{1}{N_s^3 N_\tau} \left( \left. a \frac{d \beta}{da} \right|_{\mathrm{LCP}} \left\langle \frac{\partial S}{\partial \beta} \right\rangle_{sub.}  + \left. a \frac{d \kappa}{da} \right|_{\mathrm{LCP}} \left\langle \frac{\partial S}{\partial \kappa} \right\rangle_{sub.} 
  + \left. a\frac{\partial j}{\partial a} \right|_{\mathrm{LCP}} \left\langle \frac{\partial S}{\partial j} \right\rangle_{sub.} \right).\nonumber\\ \label{eq:trace-anomaly}
\eeq
Here,  $a$ is the lattice spacing. 
We take all physical observables in the $j \rightarrow 0$ limit. Here, we eliminate the third term on the right side. 
$\langle \mathcal{O} \rangle_{sub.} (\mu) $ denotes the subtraction of the vacuum quantity. 
Thus, ideally, we should take $\langle \mathcal{O} \rangle_{sub.} (\mu) = \langle \mathcal{O} (\mu,T)  \rangle - \langle \mathcal{O} (\mu=0,T=0) \rangle $, but the simulation at absolute zero temperature is practically difficult.
In this work, we take $\langle \mathcal{O} \rangle_{sub.} (\mu) = \langle \mathcal{O} (\mu, T=79 \mathrm{~MeV})  \rangle - \langle \mathcal{O} (\mu=0, T= 79 \mathrm{~MeV}) \rangle $.
Also, utilizing the scale setting function (Eq.\ (23) in Ref.~\cite{Iida:2020emi})  and a set of $(\beta,\kappa)$ with a fixed mass ratio between pseudoscalar and vector mesons $m_{PS}/m_V$ (Table~$1$ in Ref.~\cite{Iida:2020emi}), 
we obtain the coefficients nonperturbatively as
 \beq
 a d\beta /da|_{\beta=0.80,\kappa=0.159}=-0.352, \quad a d\kappa/da |_{\beta=0.80,\kappa=0.159}=0.0282.\label{eq:beta-fn}
 \eeq


\section{Our simulation results}
\begin{figure}[htbp]
 \begin{center}
    \includegraphics[keepaspectratio, scale=0.75]{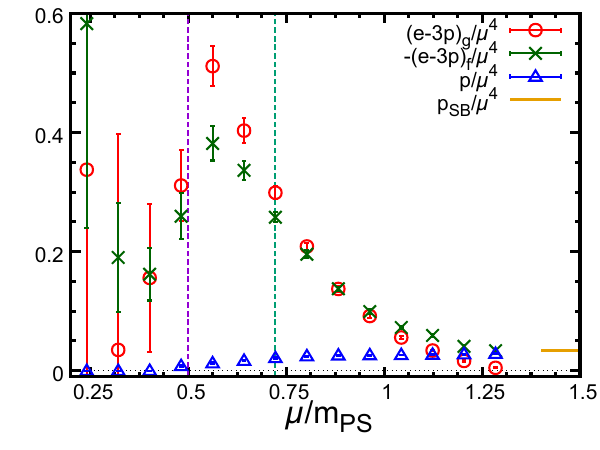}
    \caption{Trace anomaly and pressure  as a function of $\mu/m_{PS}$. The circle and cross symbols denote the gauge part and minus the fermion part of the trace anomaly, respectively.  We also show $p/\mu^4$ at the relativistic limit, $p_{SB}/\mu^4=N_fN_c /(12\pi^2)$. The plot is originally shown in Ref.~\cite{Iida:2022hyy}.
    }\label{fig:raw-data}
    \end{center}
  \end{figure}

The numerical results for the trace anomaly and pressure are shown in Fig.~\ref{fig:raw-data}.
For the trace anomaly, we plot the gauge part (the first term in Eq.~\eqref{eq:trace-anomaly}) and minus the fermion part (the second term) separately.
Both parts are normalized by $\mu^4$ to see the dimensionless asymptotic behavior.
The magnitude of each part has a peak around the hadronic-superfluid phase transition. 
It is very similar to the emergence of the peak of $(e -3p)/T^4$ around the hadronic-QGP phase transition at $\mu=0$~\cite{Borsanyi:2013bia,HotQCD:2014kol}.

As for the pressure, $p_{SB}(\mu)$ in Fig.~\ref{fig:raw-data} denotes the pressure value at the Stefan-Boltzmann (SB) limit, which is intentionally obtained by the numerical integration of the number density of quarks of the relativistic free theory to reduce the discretization effect.
In the continuum theory, the pressure scales as $p_{SB}(\mu) = \int^\mu n_{SB}^{cont.}(\mu')d\mu' \approx N_fN_c \mu^4 /(12\pi^2)$ in the high $\mu$ regime, where $N_f$ ($N_c$) is the number of flavors (colors).
We can see that our data monotonically increase after the hadron-superfluid phase transition and approaches the value of the relativistic free theory.
The value of $p/p_{SB}$ is $0.84$ at the highest density in our simulation.

 \begin{figure}[htbp]
 \begin{center}
    \begin{minipage}[b]{0.45\linewidth}
        \includegraphics[keepaspectratio, scale=0.65]{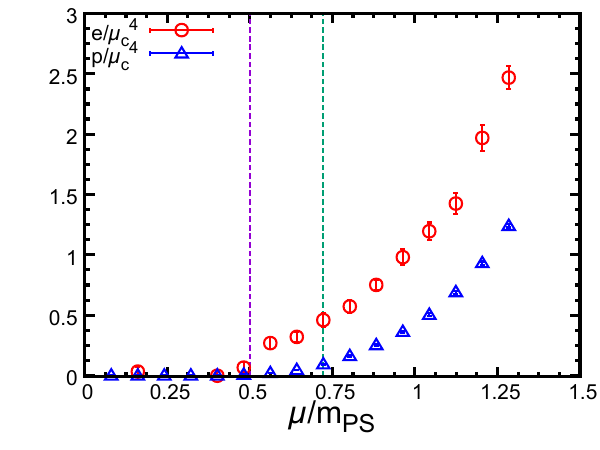}
    \end{minipage}   
    \begin{minipage}[b]{0.45\linewidth}
     \includegraphics[keepaspectratio, scale=0.65]{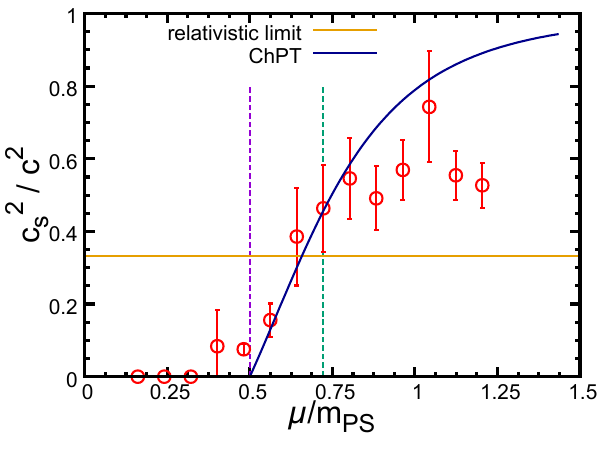}
     \end{minipage} 
    \caption{
(Left) The EoS as a function of $\mu/m_{PS}$.  (Right) Sound velocity squared as a function of $\mu/m_{PS}$. The horizontal line (orange) denotes the value in the conformal bound (the relativistic limit), $c_s^2/c^2 =1/3$. The blue curve shows the result of ChPT~\cite{Son_2001, Hands2006-mh}. These plots are originally shown in Ref.~\cite{Iida:2022hyy}.
}\label{fig:EoS}
\end{center}
  \end{figure}
Combining the data for $e-3p$ and $p$ obtained above, we finally obtain the EoS and sound velocity as depicted in Fig.~\ref{fig:EoS}.
In the left panel, we normalize $e$ and $p$ by 
$\mu_c$ so as to be dimensionless. 
We can see that both $e$ and $p$ are consistent with zero in the hadronic phase.
Thus, these thermodynamic quantities are not changed before the hadronic-superfluid phase transition even if $\mu$ increases. This is consistent with the Silver Blaze phenomenon.
Note that $e\approx 0$ in the hadronic phase indicates that the nonperturbative beta-functions for $\beta$ and $\kappa$ given by Eq.~\eqref{eq:beta-fn} work well enough to make the parts of trace anomaly, $(e -3p)_g$ and $(e -3p)_f$, cancel each other.

Finally, the sound velocity is depicted in the right panel in Fig.~\ref{fig:EoS}.
Here, we evaluate 
\beq
c_s^2 (\mu)= \frac{\Delta p (\mu)}{\Delta e (\mu)} = \frac{ p(\mu +\Delta \mu) - p(\mu -\Delta \mu)}{e(\mu +\Delta \mu) - e(\mu -\Delta \mu)}
\eeq
at the fixed temperature.
In this panel, the blue curve represents the ChPT prediction which is given by  $c_s^2/c^2=(1-\mu_c^4/\mu^4)/(1+3\mu_c^4/\mu^4)$~\cite{Son_2001, Hands2006-mh}.
The ChPT analysis is valid around the phase transition point between the hadronic and superfluid phases ($\mu \approx m_{\pi}/2$), and we can see that our lattice data is consistent with the prediction. In the high-density regime, on the other hand, the curve given by the ChPT goes to unity, namely, the sound velocity approaches the speed of light. Therefore, it is widely believed that the ChPT would fail at sufficiently high densities and that eventually the squared sound velocity goes to $1/3$.

A question to be addressed here is where it fails.
Our numerical result for $c_s^2/c^2$ is consistent with the ChPT prediction until 
the sound velocity exceeds the conformal bound.
Such an excess over the conformal bound is a characteristic feature previously unknown in any lattice calculations for QCD-like theories.
For example, at $\mu=0$ and $T>T_c$,
the sound velocity squared monotonically increases and approaches $1/3$ as the temperature increases~\cite{Borsanyi:2013bia,HotQCD:2014kol}. 

Let us comment on the definition of the sound velocity.
In our work, we calculate $\partial p /\partial e |_{T=\mathrm{const.}}$. In the standard definition, however, the sound velocity squared is given by $\partial p /\partial e |_{s=\mathrm{const.}}$ where $s$ denotes the entropy density.
Technically, it is hard to evaluate the latter one in lattice simulations, but at $T=0$ both quantities are equivalent to each other. Therefore, the estimation of the temperature dependent difference between these quantities will be one of the most important tasks in the future.

\section{Discussions}
\subsection{Still higher-density regime}
In the high-density limit, it is believed that the EoS matches with the relativistic free theory, so that $c_s^2/c^2$ should go to $1/3$.
However, it remains to be understood how it approaches $1/3$.
In the lattice simulation, we have a constraint for the value of $\mu$, namely, $a\mu \ll 1$. Otherwise, the lattice discretization error becomes sizable.
In our simulation, we take $a \mu \le 0.8$. To investigate a much higher-density regime, a finer lattice simulation or a lighter quark mass simulation where $m_{\pi}$ also becomes lighter is needed, but such a simulation is costly.

 \begin{figure}[htbp]
 \begin{center}
    \begin{tabular}{c}
        \includegraphics[keepaspectratio, scale=0.3]{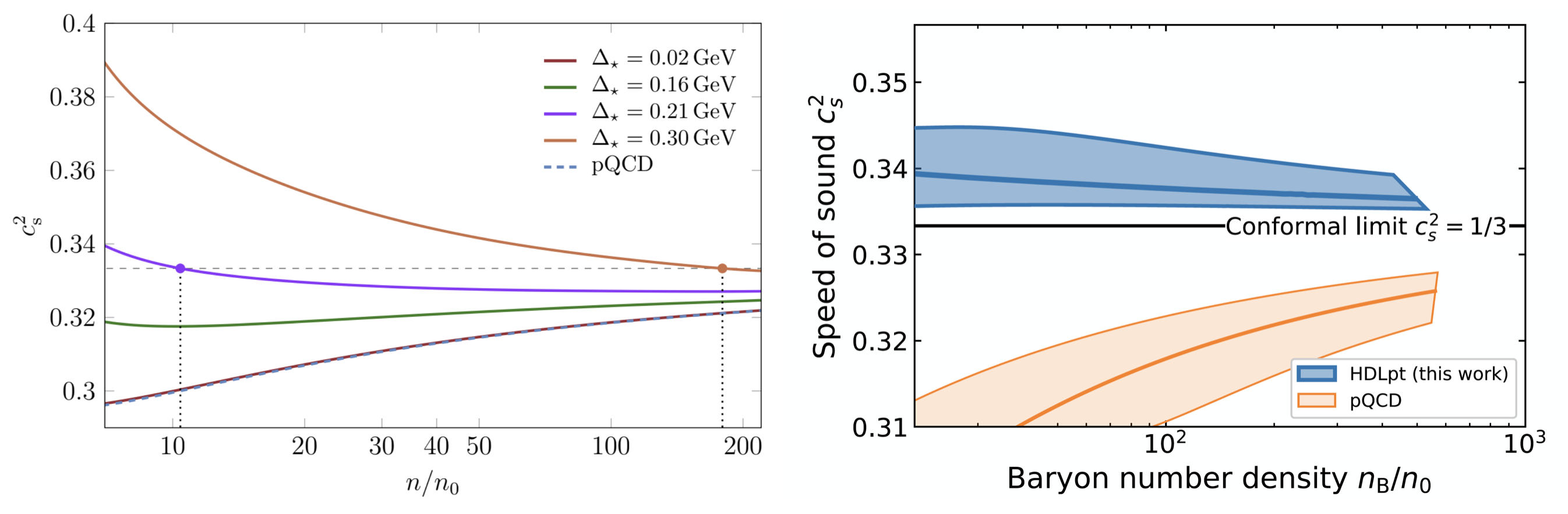}
    \end{tabular}
            \caption{
 (Left) Functional renormalization group for an effective theory of 3-color QCD with diquark gap ($\Delta$), originally given in Ref.~\cite{Braun:2022jme}.
(Right) Hard-thermal-loop resummation analysis, originally given in Ref.~\cite{Fujimoto2020-bh}. 
}\label{fig:high-density}
\end{center}
  \end{figure}
On the other hand, the pQCD analysis gives 
\beq
c_s^2/c^2 = \frac{1-5 \beta_0 \alpha_s^2/(48\pi^2)}{3},
\eeq
where $\beta_0$ and $\alpha_s$ denote the coefficient of the one-loop beta-function and the running coupling constant, respectively (see for instance Ref.~\cite{Kojo:2021wax}).
It indicates that $c_s^2/c^2$ approaches $1/3$ from below as shown in the right panel of Fig.~\ref{fig:high-density}.

On the other hand, analyses using the functional renormalization group for an effective theory of $3$-color QCD with diquark gap ($\Delta$)~\cite{Braun:2022jme} and the hard-thermal-loop resummation technique~\cite{Fujimoto2020-bh} suggest that $c_s^2/c^2$ still exceeds the conformal bound in a fairly high-density regime.

\subsection{Results for 3-color QCD and others}
Recently, other kinds of lattice results for the sound velocity at finite density have also been reported.
It is well-known that theoretical structure and low-energy phenomena of $3$-color QCD at finite isospin chemical potential ($\mu_I = \mu_u = - \mu_d $) are very similar to the case of QC$_2$D at real quark chemical potential.
In both models, a color singlet condensate emerges in a high-density regime.
In fact, $3$-color QCD at nonzero isospin chemical potential is also free from the sign problem, and at $\mu_I = m_{\pi}/2$ and at sufficiently low temperature, the pion condensation emerges.

\begin{figure}[htbp]
 \begin{center}
    \begin{tabular}{c}
        \includegraphics[keepaspectratio, scale=0.3]{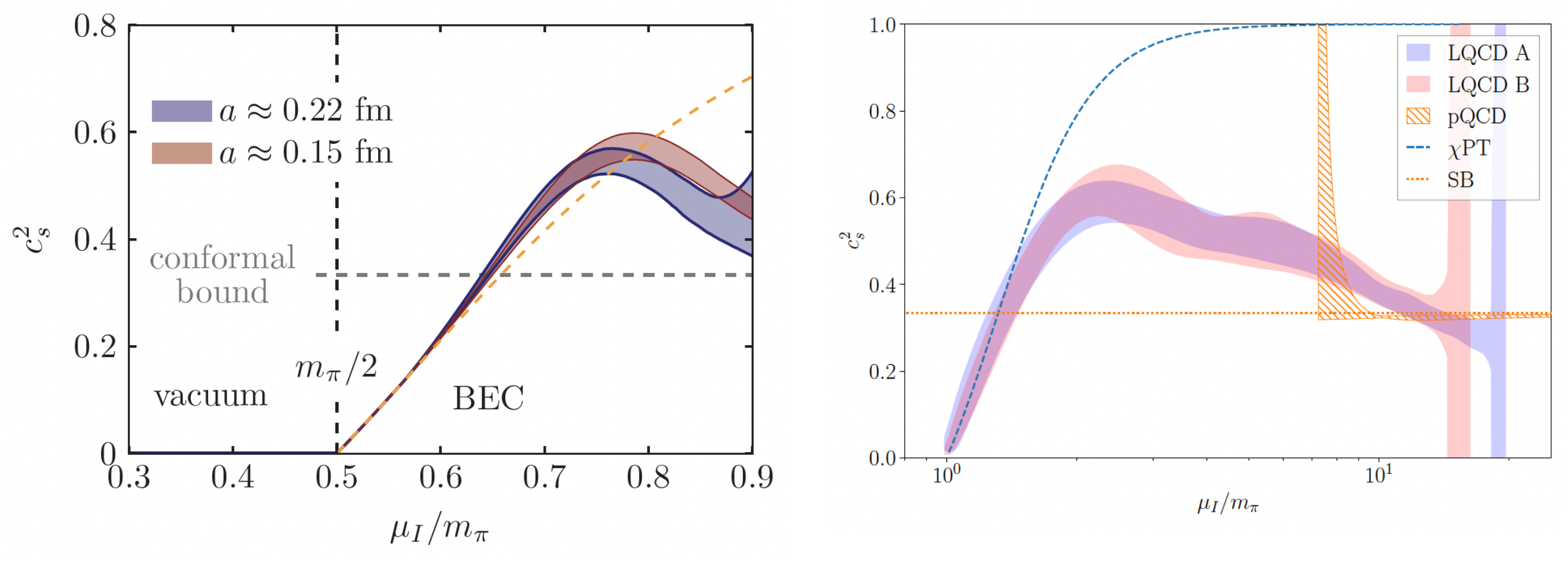}
    \end{tabular}
            \caption{
The lattice Monte Carlo results for the sound velocity squared in $3$-color QCD at finite isospin chemical potential.
The left and right panels are originally given in Refs.~\cite{Brandt:2022hwy} and \cite{Abbott:2023coj}, respectively.
}\label{fig:3-color}
\end{center}
  \end{figure}

A similar lattice Monte Carlo simulation with ours has been done by Ref.~\cite{Brandt:2017oyy}, where to solve the onset problem the pion source term is introduced instead of the diquark source term in our lattice action.
The result for the sound velocity obtained by a spline fit is depicted in the left panel of Fig.~\ref{fig:3-color}, which is originally given in Ref.~\cite{Brandt:2022hwy}.
The result is also consistent with the ChPT prediction just after the phase transition point.
Another lattice result is presented in the right panel of Fig.~\ref{fig:3-color}, which is originally given in Ref.~\cite{Abbott:2023coj}. Here, the authors estimate the sound velocity using a novel technique for calculating a multi-number of pion correlation functions. The result for the sound velocity exhibits the excesses over the conformal bound even in an extremely high-density regime. 

\begin{figure}[htbp]
 \begin{center}
    \begin{tabular}{c}
        \includegraphics[keepaspectratio, scale=0.3]{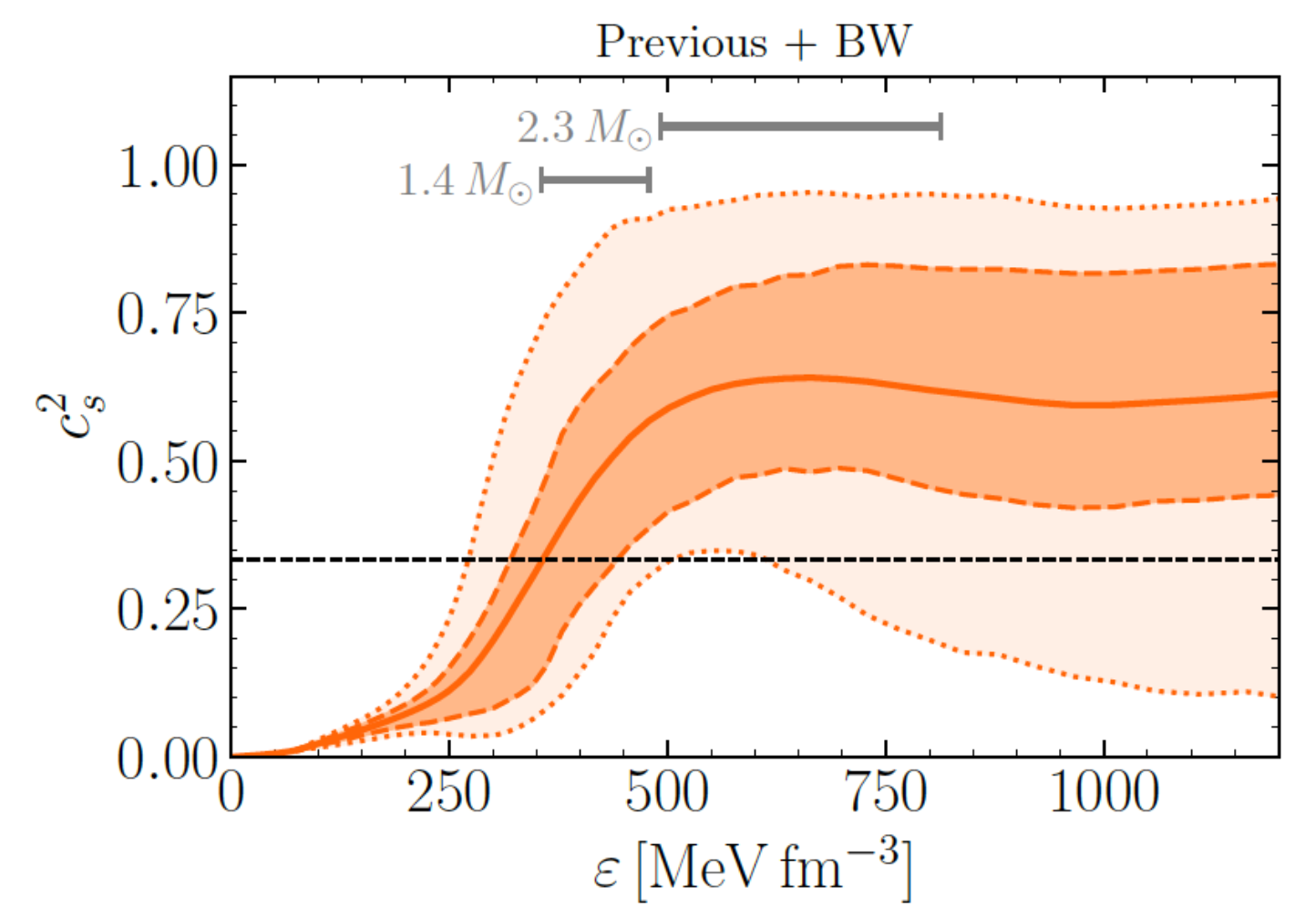}
    \end{tabular}
            \caption{
The sound velocity squared constrained from recent neutron star observational data. The plot is originally shown in Ref.~\cite{Brandes:2023hma}.
}\label{fig:NS-data}
\end{center}
  \end{figure}

The final question is whether such a breaking of the conformal bound is true even for real finite-density QCD.
Unfortunately, neither dense QC$_2$D nor $3$-color QCD at finite $\mu_I$ exactly tells us about real finite-density QCD.
The first-principles calculation for real finite-density QCD in a large box is still an extremely hard task.
On the other hand, it has been reported that recent neutron star observational data including a $2.35M_\odot$ neutron star as well as simultaneous constrains on neutron star masses and radii
also suggest an excess over the conformal bound as shown in Fig.~\ref{fig:NS-data}~\cite{Brandes:2023hma}.
The analysis indicates evidence against a strong first-order phase transition, of which the occurrence has been believed for a long time.

\section{Summary and future directions}
In our work~\cite{Iida:2022hyy}, we have exhibited the breaking of the conformal bound for dense QC$_2$D using the first-principles calculation.
It was previously unknown from any lattice calculations for QCD-like theories, but recently two other lattice results~\cite{Brandt:2022hwy, Abbott:2023coj} have also shown the breaking of the conformal bound in the context of $3$-color QCD at nonzero isospin chemical potential.  It would be fascinating to further clarity the peak of the speed of sound and its relation with the hadron-quark crossover from first-principles calculations. 

To this end, however, many questions remain.  A finer lattice simulation or a lighter quark mass simulation would enable us 
to clarify the behavior of the sound velocity at higher densities.
To make better estimates of the sound velocity in a lattice setting of nonzero temperature, furthermore, it is important to know the temperature-dependent difference between
$\partial p/\partial e|_{T=\mathrm{const.}}$
and $\partial p/\partial e|_{s=\mathrm{const.}}$.
Most interestingly, the presence of very massive neutron stars, which implies that the EoS of neutron star matter is fairly stiff at densities relevant for neutron star cores~\cite{Brandes:2023hma}, might have some relevance to effective repulsion between hadrons in medium.
It is thus essential to investigate the hadron interaction at nonzero quark/isospin chemical potential,  which will be a challenging future work~\cite{Murakami:2023ejc}.


\acknowledgments
We would like to thank S.~Hands, T.~Hatsuda,  T.~Kojo, T.~Saito, J.-I.~Skullerud, D.~Suenaga, H.~Tajima, and H.~Togashi for useful conversations.
The work of K.~I. is supported by JSPS KAKENHI with Grant Numbers 18H05406 and 23H01167.
The work of E.~I. is supported by JSPS KAKENHI with Grant Number 23H05439,
JST PRESTO Grant Number JPMJPR2113,
JSPS Grant-in-Aid for Transformative Research Areas (A) JP21H05190, 
JST Grant Number JPMJPF2221  
and also supported by Program for Promoting Researches on the Supercomputer ``Fugaku'' (Simulation for basic science: from fundamental laws of particles to creation of nuclei) and (Simulation for basic science: approaching the new quantum era), and Joint Institute for Computational Fundamental Science (JICFuS), Grant Number JPMXP1020230411.
The numerical simulation is supported by the HPCI-JHPCN System Research Project (Project ID: jh220021).
The work of E.~I is supported also by Center for Gravitational Physics and Quantum
Information (CGPQI) at YITP.

\bibliographystyle{utphys}
\bibliography{2color}

\end{document}